\documentclass[iop]{emulateapj}
\usepackage{apjfonts}
\usepackage{natbib}
\usepackage{amsfonts}
\usepackage{amsmath}
\usepackage[usenames]{color}
\usepackage{verbatim}
\newcommand\msun{M$_\odot$}
\newcommand\mzams{$M_{\rm ZAMS}$}

\newcommand{\epssclone}{\epsscale{1.0}}

\shorttitle{}
\shortauthors{}

\begin{document}

\title{The Progenitor Mass of SN 2011dh from Stellar Populations Analysis}

\author{Jeremiah W. Murphy\altaffilmark{1,2}}
\author{Zachary G. Jennings\altaffilmark{1}}
\author{Benjamin Williams\altaffilmark{1}}
\author{Julianne J. Dalcanton\altaffilmark{1}}
\author{Andrew E. Dolphin\altaffilmark{3}}

\altaffiltext{1}{Astronomy Department, The University of Washington
  Seattle, WA 98195; jmurphy@astro.washington.edu}
\altaffiltext{2}{NSF Astronomy and Astrophysics Postdoctoral Fellow}
\altaffiltext{3}{Raytheon, 1151 E. Hermans Road, Tucson, AZ 85706; adolphin@raytheon.com}

\begin{abstract}
Using Hubble Space Telescope (HST) photometry, we characterize the age
of the stellar association in the vicinity of supernova (SN) 2011dh
and use it to infer the zero-age main sequence mass (\mzams) of the
progenitor star.  We find two distinct and significant star formation
events with ages of $<$6 and 17$^{+3}_{-4}$~Myrs, and the
corresponding \mzams\ are $>$29 and 13$^{+2}_{-1}$~\msun,
respectively.  These two bursts represent 18$^{+4}_{-9}$\% (young) and
64$^{+10}_{-14}$\% (old) of the total star formation in the last
50~Myrs.  Adopting these fractions as probabilities suggests that the
most probable \mzams\ is $13^{+2}_{-1}$~\msun.  
These results are most sensitive to the
luminosity function along the well-understood main sequence and are
less sensitive to uncertain late-stage stellar evolution.
Therefore, they stand even if the progenitor suffered disruptive
post-main-sequence evolution (e.g. eruptive mass loss or binary
Roche-lobe overflow).  Progenitor identification will help to
further constrain the appropriate population.  Even though pre-explosion
images show a yellow supergiant (YSG) at the site of the SN, panchromatic SN
light curves suggest a more compact star as the progenitor.  In spite
of this, our results suggest an association between the YSG and the
SN. Not only was the star located at the SN site, but reinforcing an
association, the star's bolometric luminosity is consistent with the
final evolutionary stage of the 17~Myr old star burst. If the YSG
disappears, then \mzams$=13^{+2}_{-1}$~\msun, but if it persists, then
our results allow the possibility that the progenitor was an unseen
star of $>$29~\msun.

\end{abstract}

\keywords{supernovae: general --- supernovae: individual (SN 2011dh)}

\section{Introduction}
\label{section:introduction}

Observational measurements of the masses of supernova (SN) progenitors
are currently scant and highly uncertain.  Of the $\sim$20 SNe that
have progenitor mass constraints, only about half have well-defined masses
and the rest have upper bounds \citep{arnett89,aldering94,barth96,
vandyk99,vandyk02,smartt02,vandyk03a,vandyk03b, smartt04,maund05,
hendry06,li05,li06,li07,galyam07,crockett08,galyam09,smartt09a,
smartt09b,smith11b,smith11c}. In general, stellar evolution theory
predicts a clear mapping between the zero-age main sequence mass
(\mzams) and the explosion scenario for isolated massive stars
\citep{woosley02,heger03a,dessart11}.  
These limited observations suggest that the least
massive stars explode as SN II-P as expected \citep{smartt09a}. More
massive stars are expected to lose much of their
mass and explode as H-deficient SNe (IIb and Ib/c).
However, some H-rich SNe (in particular IIn) have been
associated with very massive stars \citep{galyam09,smith11b,smith11c}.
Furthermore, theory \citep{claeys11,dessart11} and the relatively high
observed rates of H-deficient SNe
\citep{smith11c} imply that binary evolution may figure prominently in
producing the H-deficient SNe. Even in light of these complications, the maximum
\mzams\ associated with SN II-P is much lower than expected
\citep{smartt09a}.  While tantalizing, these initial results are
poorly constrained, and even the simple statement that stars more
massive than $\sim$8~\msun\ explode as SNe requires more observational
constraints.  Thus every new supernova marks an important opportunity
to add a new progenitor mass measurement to this poorly-constrained sample.

Recently, amateurs and the Palomar Transient Factory
(PTF) collaboration detected supernova (SN) 2011dh
in M51, the Whirlpool Galaxy \citep{griga11,silverman11,arcavi11b}.  
Initial reports classified it as
II \citep{silverman11}, but late time spectra reveals this explosion to
belong to the rare IIb transitional class \citep{arcavi11b,marion11},
indicating that the progenitor lost much of its hydrogen envelope due
to either binary evolution or unknown single-star mass-loss physics.
The close proximity of this SN \citep[$(M-m)_0=29.42$, 7.7
Mpc;][companion galaxy NGC~5195]{tonry01} and the wealth of Hubble
Space Telescope (HST) data on M51 offers a rare opportunity to further
explore the physics of the progenitor.

Using archival HST photometry, \citet{maund11} and
\citet{vandyk11} identified a yellow supergiant (YSG) as
the progenitor candidate.  \citet{vandyk11} fit the magnitude and color
of the progenitor to evolutionary tracks and derive a
zero-age-main-sequence progenitor mass of \mzams = 18-21~\msun.
However, \citet{maund11} argue that an uncertain mass-loss history
causes the color to be an unreliable characteristic of the progenitor.
Instead, they treat the bolometric luminosity as an intrinsic property
determined by the core mass, which is in turn determined by
\mzams.  Matching the
bolometric luminosity to the very last stages of evolutionary models,
they derive \mzams = 13$\pm 3$~\msun.  To further
complicate the situation, recent observations suggest that the YSG may not even be the progenitor.  Based upon the
characteristics of the SN light curves,
\citet{arcavi11b} and \citet{soderberg11} argue that the YSG is not the
progenitor and instead suggest a more compact source.

This discrepancy in mass measurements highlights the fact that
interpretation of the precursor photometry alone is sensitive to
uncertain mass-loss physics in the final evolutionary stages of the
models.  Furthermore, using precursor imaging to measure progenitor
masses requires that precursor imaging exists and that the SN position
is known to sub-arcsecond precision.  The majority of past SNe have
neither pre-existing HST imaging, nor sufficiently accurate
astrometry.  A significant fraction of future nearby SNe will lack
precursor imaging as well.

In this letter, we address the progenitor mass with an independent
approach; we characterize the age of the stellar association
surrounding SN~2011dh and derive the corresponding \mzams\ for the
progenitor.  This technique was thoroughly described by \citet{gogarten09} on a
SN ``impostor'' in NGC~300 and by \citet{badenes09} on SN remnants in
the Magellanic Clouds.  In brief, using well-established stellar
population modeling techniques, we can age-date the star formation
episode that led to the observed SN.  The resulting age can place
strong constraints on the mass of the precursor by leveraging the
well-understood properties of a large number of main-sequence
stars. This complementary approach is feasible even when there is no
precursor imaging, or when the SN position is only localized to within
several arcseconds.

In \S2, we detail our application of this technique to SN~2011dh, the
most distant and only ongoing SN to which it has been applied.  In \S3, we
provide the resulting age and mass measurements, and in \S4, we
compare our resulting progenitor mass with those obtained using direct
imaging of the progenitor \citep{maund11,vandyk11}.  We find our
results to be most consistent with the results of \citep{maund11}
and improve upon their uncertainties.

\section{Data \& Analysis}
\label{section:data}

\subsection{Data Acquisition and Photometry}

The region surrounding SN~2011dh
was imaged with HST using the Advanced Camera for Surveys (ACS) on
January 20-21, 2005 (Hubble Heritage Team, target name
M51-POS5, Proposal 10452).  For our analysis, we retrieved
calibrated data from the HST Archive in the deepest two of the four
available filters, F555W and F814W.  Four dithered images of exposure
time 340 s were available in each filter, providing a total exposure
time of 1360 s in each band.

Resolved stellar photometry was performed using the photometry
pipeline developed for the ACS Nearby Galaxy Treasury program
\citep{dalcanton2009}.  This pipeline uses the DOLPHOT stellar
photometry package \citep{dolphin2000} to fit the ACS point spread
function (PSF) to all of the point sources in the images.  Fluxes are
then converted to Vega magnitudes using the standard zero-points and aperture corrections from the ACS handbook.
To assess photometric errors and completeness, at least 10$^5$ fake
star tests are performed by inserting fake stars of known color and
magnitude into the data 
and blindly attempting to
recover them with the same software.  For details of the quality cuts,
see \citet{dalcanton2009}.

\subsection{Recent Star Formation History Recovery}

For our analysis, we selected stars within 1.4$''$ ($\sim$50~pc) of the SN
location \citep{gogarten09}.  By varying the radius from 30 to 100 pc, we verify
that 50~pc maximizes the SF in the local bursts and minimizes contamination.
This resulted in 29 stars being selected,
including 16 upper main sequence stars. A color-magnitude diagram
(CMD) of these stars is displayed in the left panel of
Fig.~\ref{cmdplots}.

We fit the stellar evolution models of \citet{girardi2002,girardi10}
to our data using the software package MATCH, which calculates a
maximum likelihood fit of a linear combination of model CMDs to the
observed CMD \citep{dolphin02}.  We verified that our CMD was
consistent with a population located at the published distance of M51
\citep[$(m-M)_0$=29.42,][]{tonry01} by allowing MATCH to fit the
distance modulus. The best-fit distance modulus is
$(m-M)_0$=29.45 and is consistent with the published distance.  
MATCH also fits for a distribution of extinctions; we fix the width of
the distribution at 0.5, but allow the minimum to vary from Av=0.0 to 0.5.
We constrained the metallicity to increase for younger ages.
We only fit data brighter than our 50\% completeness limit as measured
by our artificial star tests.  These limits were $m\,<$26.95 in both
filters.

Having calculated a best-fit age distribution, we quantify the random
and systematic uncertainties.
The random error is due
to sampling statistics and is largely determined by the
number of stars in the sample.  The systematic error is due to
potential offsets between model and observed CMDs.  For example, if the model is systematically
slightly bluer than the data (potentially due to systematic offsets in
the model transformations or model deficiencies), then the resulting
age distribution will be affected.  We estimate the uncertainty from
both of these sources through a series of Monte Carlo (MC)
realizations of the data.  Each realization draws the same number of
stars as our data from the best-fit age distribution.  These random
draws are re-fit with the models after shifting the models by
0.4 magnitudes in bolometric luminosity and 7\% in
effective temperature, which represent distance
  and systematic model uncertainties.  Thus, the set of resulting fits provide a
distribution of measurements from which we calculate our uncertainty.
Finally, to assess the dependence of the result on individual
outliers, we performed jackknifing tests in which the brightest or
bluest star in the field (such as the SN progenitor candidate) was
removed from the data.

Our goal in this analysis is to identify the burst of star formation
in which the progenitor star was most likely created.  We therefore
focus on star formation history (SFH) of the past 50~Myr.
To identify recent star formation bursts, we examine the cumulative star
formation as a fraction of the total stellar mass formed in the past
50~Myr.  We then calculate the median age of the recently formed stars
as well as the width of their age distribution.  This method has the
advantage of only being sensitive to relative star formation rates,
not absolute values.  Furthermore, the cumulative distribution is less
affected by the covariances between neighboring age bins in the model
fits.

This methodology leads to two sources of uncertainty in
age. 
One is the
uncertainty in the median age, which the MC tests address, and the
second is the width of the distribution of ages.
To account for the uncertainty
due to the intrinsic age, we assign probabilities to each age based on
the mass fraction of stars with that age.  Uncertainties on these
probabilities then account for the measurement errors.  Our quoted
mass values correspond to the most massive star for the age limits,
which are calculated directly from the same models as used for fitting
\citep{girardi10}.

\section{Results}
\label{section:results}

\begin{figure}
\epssclone
\plottwo{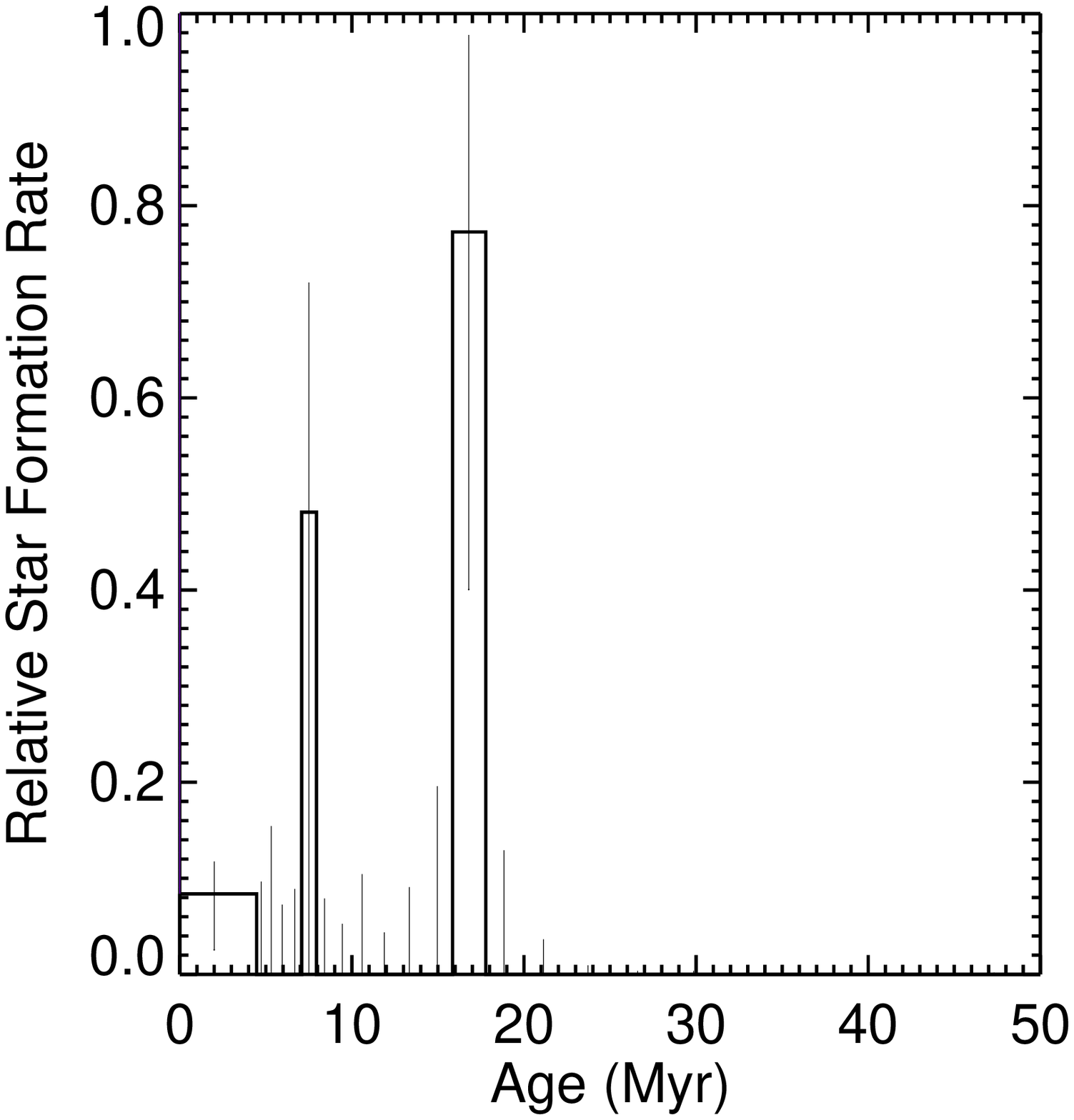}{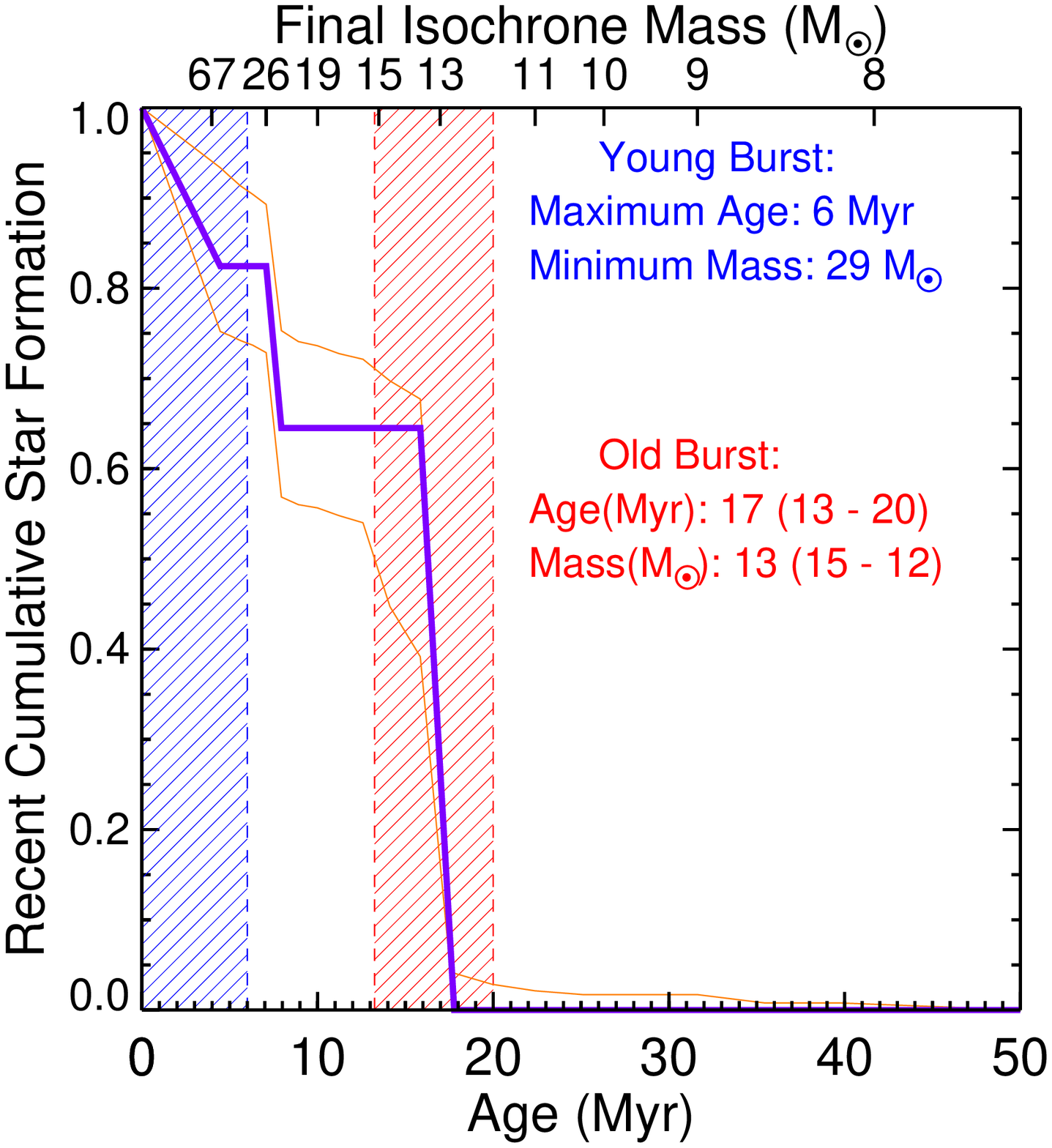}
\caption{Modeled star formation history (SFH) in the vicinity of SN
  2011dh.  The left panel shows the relative star formation rate
  (solid black line).  We use
  the cumulative SF since 50~Myr ago (right panel) to characterize the
  age.  We find two significant SF events at $<$6~Myr and
  17$^{+3}_{-4}$~Myr  (See \S~\ref{section:conclusions} for a
  discussion of how we rule out the 8~Myr old burst).  The
  corresponding \mzams\ for the progenitor are $>$29 and 13$^{+2}_{-1}$~\msun.  Most of the recent SF occurred in the 17~Myr burst.
  Therefore, our most probable \mzams\ estimate for the progenitor is 13$^{+2}_{-1}$~\msun.\label{sfhplots}}
\epsscale{1.0}
\end{figure}

In this section, we report the best-fit metallicity, average
extinction, and SFH of the region surrounding the location of SN
2011dh.  In Fig.~\ref{sfhplots}, we show the primary result, the age
distribution of young populations surrounding SN~2011dh.  The best-fit
metallicity for the last 50~Myr is solar, which compares favorably
with the range of metallicities derived from HII regions in M51
\citep{Bresolin04}.  For the average extinction, we find $A_V =
0.425$.  The value for galactic extinction at this pointing is
$A_V$=0.12 \citep{schlegel98}; therefore this region likely contains
additional extinction from dust within M51. To estimate the
uncertainty due to extinction, we force $A_v$=0.12 in MATCH and
compare these results to the self-consistent fit.  The resulting
differences in ages and masses are included in the uncertainties that
we report in this letter.

\begin{figure}
\epssclone
\plottwo{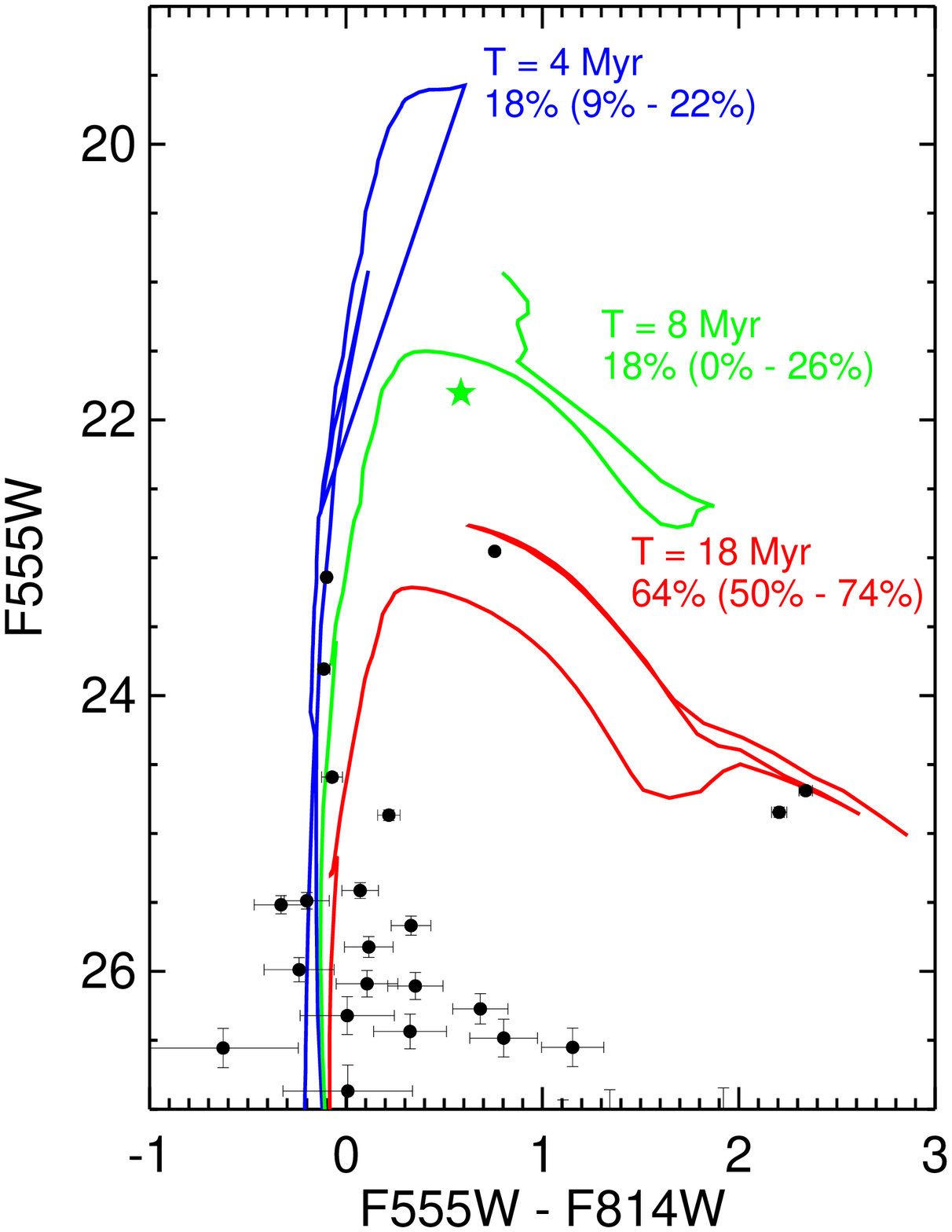}{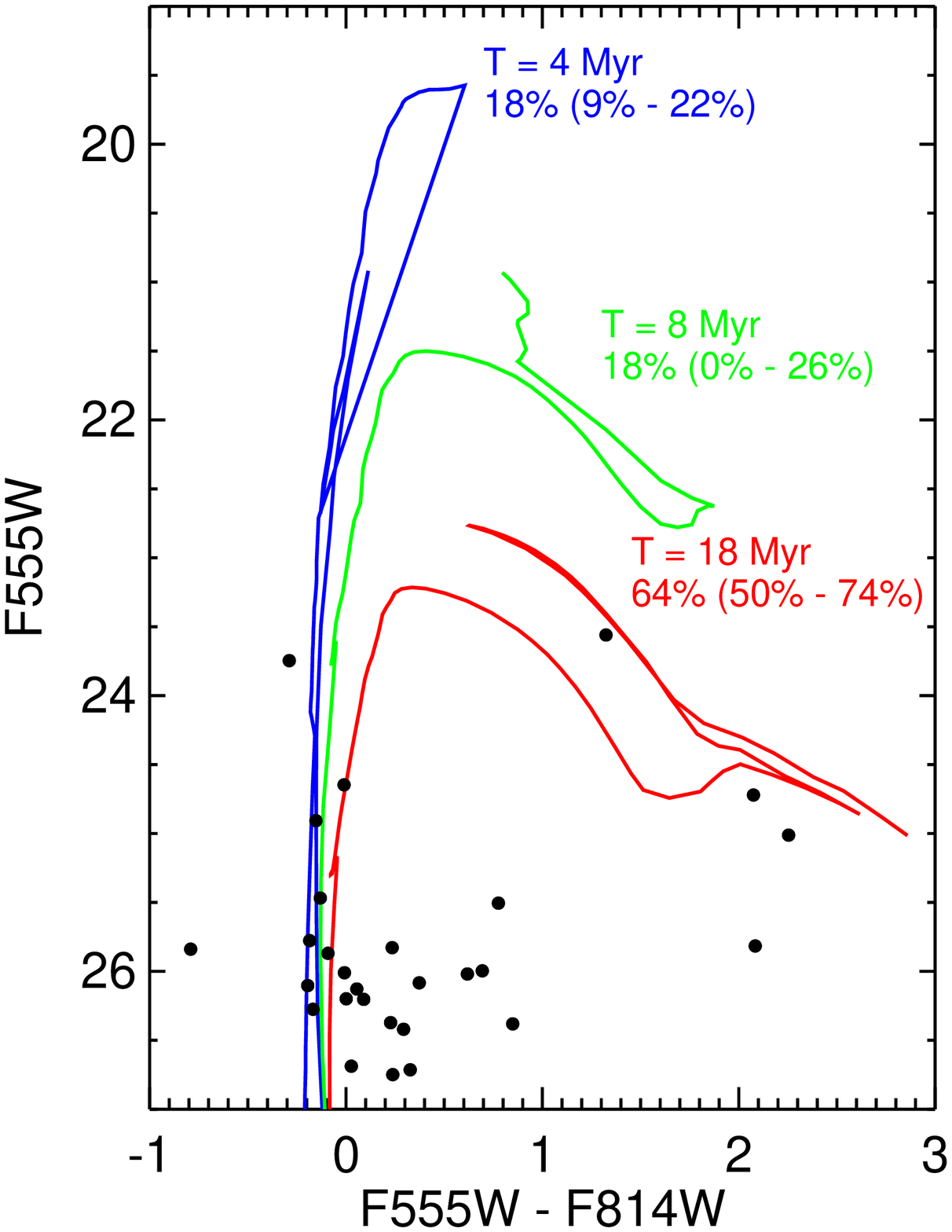}
\caption{Observed (left panel) and modeled (right panel)
  color-magnitude diagrams (CMD).  In the left panel, the brightest
  star (green star symbol) is the progenitor candidate
  \citep{maund11,vandyk11}.  
The three lines correspond to Padova isochrones that are
  closest in age to the three SF bursts in Fig.~\ref{sfhplots} and each is labeled with the age
  and percent of total SF in the last 50~Myr.  Note that we do not fit these
  isochrones to determine the age.  Rather, we model the CMD and find
  the SFH that best represents the data. The right panel shows a
  representative CMD model from the MC simulations. 
\label{cmdplots}}
\epsscale{1.0}
\end{figure}

The left panel of Fig.~\ref{cmdplots} shows the magnitude (F555W) and
color (F555W-F814W) of the brightest stars surrounding the SN, which
we use to calculate the SFH in the vicinity of SN 2011dh. The green
star shows the magnitude and color of the progenitor candidate, a
YSG \citep{maund11,vandyk11}.  For comparison, we plot
three isochrones corresponding to the three star formation (SF) events in
Fig.~\ref{sfhplots}.  In addition, we show the corresponding age and
fraction of total SF in the last 50~Myrs.  The right panel of
Fig.~\ref{cmdplots} shows a model CMD produced by MATCH for the
best-fit SFH, with the same isochrones over-plotted.  Note that the
model generically reproduces the number density of stars on the MS and
red giant regions of the CMD.

To be clear, we did not fit the main sequence turn-off (MTO) to arrive at
these isochrones. Because of the small number of main sequence stars at
the bright end, MTO fitting is susceptible to large Poisson errors.
Rather, as we explain in \S~\ref{section:data}, we model the entire
CMD and constrain the SFH that is required to reproduce the observed
CMD.  This way the number density of stars at each color and
magnitude, particularly along the entire MS, is involved in
constraining the age of the star burst, not just the location of the
end of the MS.

The best fit SFH is shown in Fig.~\ref{sfhplots}.  The left panel
shows the relative SFH (black solid line) as a function of age, and
the right panel shows the cumulative SF (thick solid purple line)
since 50~Myr ago.  The orange thin lines show the 68\% confidence
interval for the cumulative distribution.  In both plots, three bursts
of SF are apparent: 4, 8, and 17~Myr.  
The 17~Myr burst accounts for most (64$^{+10}_{-14}$\%) of the recent
SF in the last 50~Myr.

Based on these bursts, we report two estimates for the age and \mzams.  The age is either $<$6 or 17$^{+3}_{-4}$~Myr,
corresponding to \mzams\ of either $>$29 or 13$^{+2}_{-1}$~\msun,
respectively.  Since most of the SF in the last 50~Myr is associated
with the $\sim$17~Myr burst, we find the most probable \mzams\ of the
progenitor to be 13$^{+2}_{-1}$~\msun.  We rule out the $\sim$8~Myr SF event
for three inter-related reasons.  For one, this event has
uncertainties consistent with zero star formation.  Secondly,
jackknifing tests show that this event is strongly dependent on the
presence of the YSG, suggesting that no other stars in
the sample are consistent with this age.  Thirdly, if the YSG is the progenitor, then its color implies that it
experienced a great deal of mass-loss that is not included in the
models used in MATCH.  Hence, any fits that rely on the magnitude and
color of this star alone are in error.  See
\S~\ref{section:conclusions} for further discussion on these last two
points.  In the cumulative plot, the age and mass estimate associated
with the younger event is highlighted by the blue shaded regions,
while the older and most probable estimate is highlighted by the red
shaded region.

\section{Discussion \& Conclusions}
\label{section:conclusions}

Using HST photometry, we characterize the age of the stars
in the vicinity of SN 2011dh, and from this age, we infer
the \mzams\ of the progenitor star.  The recent SFH shows
two SF events with ages of $<$6 and 17$^{+3}_{-4}$~Myrs, corresponding to
\mzams\ for the progenitor of $>$29 and 13$^{+2}_{-1}$~\msun, respectively.  Given
that the older burst at 17$^{+3}_{-4}$~Myr represents 64$^{+10}_{-14}$\%
of the total SF in the last 50~Myrs, the most probable \mzams\ for the
progenitor is 13$^{+2}_{-1}$~\msun.  Because MATCH leverages the
entire CMD, including the MS, these \mzams estimates are relatively
insensitive to the individual peculiarities of post-MS or binary
evolution.

Stellar population analysis shows that the colors of the YSG are
consistent with the surrounding stars only if it experienced peculiar
evolution.  Testing the robustness of the three SF events, we
perform several jackknifing tests, removing a bright star and
finding new best-fit SFHs.  The $<$6 and 17~Myr old populations are robust
to these tests, but upon removing the YSG, the 8~Myr old burst
vanishes ($<$3.5\% of the recent SF).  Hence, the 8~Myr feature is solely dependent upon the color
and magnitude of the most poorly modeled star in the field, the YSG --
an unlikely scenario.  More likely, the YSG was born in one of the
other events and experienced disruptive mass loss, giving it peculiar colors.
In fact, its bolometric luminosity is consistent with the final
evolutionary stage of the 17~Myr old population.


Regardless of an association with the SN, the luminosity of the YSG \citep{maund11,vandyk11} suggests that it is associated with the 17~Myr population,
in which case, we derive a \mzams\ for the YSG of
13$^{+2}_{-1}$~\msun.  The previous \mzams\ estimates were obtained by
comparing photometry of the YSG with stellar evolution
models.  Of the two previous attempts to estimate \mzams\, one at
13$\pm$3~\msun\ \citep{maund11} and one at 18-21~\msun\
\citep{vandyk11}, the lower mass estimate is most consistent with our
results.  This consistency leads to a couple of conclusions.

For one, this consistency validates using nearby stellar ages to
estimate the progenitor mass when an archival HST image of the
progenitor is not available.  Given the distance to M51, $\sim$7.7~Mpc, the fact that this technique is able to further constrain the
\mzams\ estimate is remarkable.  Directly modeling the progenitor in
precursor imaging only requires accurate photometry for the most
luminous stars and has become a routine exercise up to 20~Mpc
\citep{smartt09b}.  Hence, for direct imaging of the precursor,
SN~2011dh is quite close.  Our technique, on the other hand, relies on
having a sufficient number of main sequence stars to characterize the
age of the associated star burst. With a distance of $\sim$7.7~Mpc,
the magnitude limit is quite high, resulting in only 16 detectable
upper main sequence stars within 50 pc of the SN.  Even with this
small number of upper main sequence stars, we were able to improve
upon the direct imaging mass constraint for the YSG.

The second conclusion we draw from the consistency is that fitting the
bolometric luminosity of the progenitor to the last stages of
evolutionary tracks \citep{maund11} is a more robust method to
estimate \mzams\ than fitting both the luminosity and effective
temperature \citep{vandyk11}.  The literature is full of theoretical
arguments suggesting that the bolometric luminosity during the last
evolutionary stages is a more intrinsic property of \mzams\
\citep{arnett89,woosley02,smartt09b}, but as far as we know, there
have been no independent observations to support this.  Because
age-dating the nearby stars leverages information from the MS phase,
our results provide a complimentary estimate of \mzams\ and
independent support for using the bolometric luminosity only in direct
imaging techniques.

While the most probable \mzams\ of the SN progenitor is 13$^{+2}_{-1}$~\msun,
there is a small but not insignificant probability that the progenitor
mass is $>$29~\msun.  Ruling out one or the other as the progenitor
requires further information.  
If the YSG persists in post-SN imaging, then the
progenitor was indeed compact \citep{arcavi11b,soderberg11}, and neither mass
estimate is ruled out.
However, if the YSG disappears,
then our progenitor mass estimate of 13$^{+2}_{-1}$~\msun\ would
validate and further constrain the \citet{maund11} estimate.
Given the YSG's positional coincidence
with the SN and the fact that its bolometric luminosity is consistent
with the final evolutionary stage of the 17~Myr old population, it
would be odd if the YSG is not associated with the SN.

\acknowledgments

We thank George Wallerstein for drawing our attention to this SN.
J.W.M. is supported by an NSF Astronomy and Astrophysics Postdoctoral
Fellowship under award AST-0802315.  Z.G.J. is supported by the same award.  This work is based on observations made with
the NASA/ESA Hubble Space Telescope, obtained from the data archive at
the Space Telescope Science Institute. STScI is operated by the
Association of Universities for Research in Astronomy, Inc. under NASA
contract NAS 5-26555.

\bibliographystyle{apj}


\clearpage

\end{document}